\documentclass[twocolumn,showpacs,preprintnumbers,amsmath,amssymb]{revtex4}

\usepackage{graphicx}
\usepackage{dcolumn}
\usepackage{bm}
\usepackage{color}
\usepackage{amsmath}
\usepackage{amsfonts}
\usepackage{amssymb}
\usepackage{natbib}
\usepackage{latexsym}
\usepackage{float}

\begin{document}
\title{ A Unified theory  of transport barriers (TBs) in magnetically confined systems }
\author{Swadesh.M. Mahajan$^1$, David Hatch$^1$,   Zensho Yoshida$^2$,and Mike Kotschenreuther$^3$ }
\affiliation{
$^1$Institute for Fusion Studies, The University of Texas at Austin,
Austin, Texas 78712, U.S.A.
\\
$^2$ Department of Mathematics, The University of Tokyo, Japan,
 \\
$^3$ ExoFusion, Bellevue, Washington-----.
}\date{\today}

\begin{abstract}

A thermodynamic model of a plasma boundary layer, characterized by enhanced temperature contrasts is proposed. The theory is constructed to determine the inner boundary temperature $T_1$ for a specified outer (colder) boundary temperature $T_0$, the heat flux $F$ entering the inner boundary, and the parameters defining the layer. The system shows bifurcation and switches to a stable high gradient state if the heat flux $F$ entering through the inner boundary exceeds a critical value $F_c$. However there is an additional stringent condition for the transition to occur; the edge temperature $T_0$ must exceed a critical value $T_c$- no transition is possible if $T_0<T_c$ even for arbitrary large $F$. Equally important is the finding that $F_c$ is not a monotonic function of $T_0$ but has a minimum at  $T_{optimum}$ (= $4T_c$ )in the model calculation. The confinement peaks at $T_{optimum}$. The basic conceptual physics is obviously simple: The high contrast state becomes the preferred state when the incoming power into the layer is preferentially converted into coherent motions like the fluid flows and currents (undermining the standard diffusive processes that keep the lower temperature contrast). The purely macroscopic thermodynamic model bears excellent comparison with   experimental and detailed microscopic investigations of the H-mode. Deeper plausibility reasons for the workability of this heat engine, creating the simultaneous existence of an ordered state and large entropy production, are suggested.

 \end{abstract}

\maketitle
 

\section{Introduction}
 This paper is an attempt to construct a  comprehensive/encompassing physical picture of the spectacular and highly important  phenomenon of transport barriers (TB) that have been successfully created  in the most advanced magnetic fusion devices of today (tokamaks and spherical tokamaks). In the world of Fusion (via Magnetic Confinement), plasma configurations with TBs, whether internal (ITB) or edge (H-Modes)~\cite{tokamak1,tokamak2}, are clearly the uniquely favored (if not the only) vehicles to commercially viable fusion power. 

Expected to play a vital role in the realization of fusion power, there is a tremendous body of experimental literature, in particular, on H-mode barriers. In the last 10-15 years, advanced semi-analytical methods and powerful modern simulation codes (spanning a variety of Magnetohydrodynamic to Gyrokinetic codes) have been harnessed to expose the ontogenic development and sustenance of H-modes. One of the most important result of these investigations has been to identify plasma processes (instabilities, for instance) that need to be suppressed (weakened) in order that patently ``non-thermodynamic" states with large temperature gradients could emerge and be sustained; detailed simulations have also been quite successful in identifying plasma regimes (characteristics of magnetic geometry, edge conditions, relative density and temperature gradients) that are more conducive to harboring TBs. This entire realm of theoretical/computational  study will be labelled as Microscopic TB (MTB) approach. Despite its success in explaining experiments, MTB may constitute only one of the building blocks of a more complete theory.

Considerably before the intensification of efforts in MTB investigations (to which this team of authors has made crucial contributions ~\cite{JETMTM,PedTransport,JETILW,FingurePrints,JETmagnetic,GloMTM,KLMHM}, a subset of us had investigated this general phenomenon within the rubric of non-equilibrium thermodynamics of macroscopic systems ~\cite{YM2008}. In this field, two seemingly opposite entropy-based ``principles'' have worked remarkably well: 1) the principle of {\it minimum entropy production} has been highly successful in predicting a variety of self-organized structures in
nonlinear systems~\cite{Prigogine}, while  2)  the principle of {\it maximum entropy production}, first proposed by Paltridge and exploited by him and others {\cite{Paltridge1979, {Wyant,Grassl,Shimozawa},Lorenz2001}}, provides the basis for how some fluid systems  may maximize entropy production with simultaneously enhancing temperature inhomogeneity. 

The rationale for why the {\it maximum entropy production} principle may lead to the creation of states exhibiting enhanced  temperature inhomogeneity was further strengthened by Ozawa {\it et al.}~\cite{Ozawa2001} who pointed out that the maximum entropy production may be a general consequence of fluid-mechanical instabilities (such as B\'enard convection or Kelvin-Helmholtz instabilities) that can work as a heat engine. In this connection, it must be mentioned that ~\cite{Dewar} developed a statistical-
mechanical model of nonequilibrium flux-driven systems, where the maximum entropy production is related to the most probable “paths” of transitions.

This demonstration of Paltridge (bolstered by the theoretical underpinnings) is of particular relevance to our enquiry since a transport barrier is nothing but  a state of {\it "enhanced  temperature inhomogeneity"}. Can a similar heat engine operate in a confined hot thermonuclear plasma? The affirmative answer to this question was worked out in some detail in ~\cite{YM2008} (to be referred to as YM08).  

The framework developed in YM08 attempted to bring into the preceding general fold the spectacular self-organization manifested in the high-confinement TBs that are routinely engineered in tokamak discharges~\cite{tokamak1,tokamak2}; the gross features of phenomena of this genre could, indeed, be ``predicted'' and explained by the principle of {\it maximum entropy production}.

The model framework was constructed by probing into the thermodynamics of a generic  plasma boundary layer (region with large gradients) in which a specified heat flux enters form the left while the right boundary is kept at a fixed temperature $T_0$ by a heat bath (Fig.~1). {\it It was shown that, under suitable conditions, the system does exhibits bifurcation; two distinct stable states of temperature distribution are possible. The total heat flux is the controlling parameter; when greater than a critical value, it tends to favor the state with a larger temperature contrast.} 

Notice that, in contrast to MTB, this macroscopic thermodynamic route (to be called Thermodynamic TB ($T_h$TB)), does not solve equations of plasma electrodynamics, but deals with very gross features that define a thermodynamic equilibrium that is blind to the details of what actual processes are taking place in the boundary layer ~ \cite{N1}. YM08, however, does proffer a set of feasibility arguments; in particular, about how such purely thermodynamic considerations (devoid of electromagnetism) could be relevant to H-modes and ITB's in Tokamak plasmas.

Referring the reader to YM08 for details of the model, we will reproduce here the most relevant mathematics, and a discussion of the predictions. The $T_h$TB and MTB are complementary approaches to the grand problem at hand; by comparing and combining them, we will have a much more complete theory to guide us not only to understand but also to engineer most effective plasma confinement systems. To the best of our knowledge, this is first time such an attempt is being made!

\section{Thermodynamics, Entropy production, Temperature contrast }

Before working out the thermodynamics of the layer, it is necessary to clearly define our objectives. We are seeking the emergence of a state that supports enhanced temperature gradients- a large $\Delta T$ across the layer. One must immediately confront the question - larger than what? Fortunately, the answer is quite straightforward- We are seeking a $\Delta T$ greater (much greater) than a base  $\Delta T_D$ sustained by the diffusive processes (both classical and turbulent) that are ambient to the layer. {\it Our success will depend on whether the inflow of energy at the inner layer can unleash processes - like the creation of flows and currents in the plasma layer - that can tamper/overcome the temperature equalizing effects of diffusive processes.} We emphasize that the diffusion suppressing processes in thermonuclear plasmas  could be driven both by flows and currents as compared to the flow-only processes available to neutral fluids considered in Ozawa {\it et al.}~\cite{Ozawa2001}- Perhaps, a plasma could turn out to be more efficient in creating-sustaining temperature inhomogeneities. For linguistic simplicity, we will use the term {\it Flows} for flows plus currents.

For magnetized plasmas, there is another observable non-diffusive "motion"- the macroscopic magnetic field; the latter could be generated by the dynamo action from plasma turbulence.

The thermodynamic preliminaries, i.e, calculating the power available to generate {\it Flows} that will allow 
$\Delta T> \Delta T_D$ are worked out in YM08. The plasma layer is idealized in which a heat flux $F_1$ enters from the inner boundary ($\Gamma_1$, adjacent to the core) while the outer boundary $\Gamma_0$ is maintained at a constant temperature $T_0$.  What is specified is the heat flux $F_1$  and temperature $T_0$ of the heat bath- what is to be determined is the inner-boundary temperature $T_1$ whose value measures the layer temperature gradient.

Since the details of this derivation are not necessary for clarity and self-sufficiency of this paper, we are putting it in Appendix A. The principal result (derived in Appendix A) that we will use in constructing the model of layer transport is contained in the equation
\begin{equation}
\int \dot{W} dM 
= \left( 1- \frac{T_0}{T_1} \right)F_1
-T_0 \int \left( \dot{S}_D+\dot{S}_i \right)dM .
\label{workdonetranstext}
\end{equation}
relating the integrated work done to the temperatures $T_1$, $T_0$  and the flux $F_1$. In Eq.(\ref{workdonetranstext}), the second term on the right hand side denotes entropy generation. The integration variable is the mass element $dM= \rho d^{3}x$. This equation will be invoked in the next section.


\section{Model of heat transport}

We will restate the profound but mathematically trivial statement that in order to maximize temperature inhomogeneity, the system must evolve a mechanism that can fight and overcome processes like heat diffusion, precisely the processes that tend to minimize the temperature inhomogeneity. The tokamak H-mode experiments are a prime example of this verity where edge transport barriers appear along with a strong sheared flow. In other words the experiments reflect the system's ability to impart energy to the collective plasma motions initiating processes that suppress diffusion; large gradients follow. We will, hopefully, demonstrate, that so many general features of the H-mode phenomenology can be understood in terms of the generic thermodynamic model we are about to recapitulate and complete.  



We begin with a simple transport model equation in which the temperature difference between 
the inner and the outer boundary is controlled by a {\it Flow} dependent heat "diffusivity",
 \begin{equation}
T_1(P) - T_0 = \eta({P}) F,
\label{Fick}
\end{equation}
where $T_1(P)$ , the inner boundary temperature we are seeking, is a function of the driving power ${P}$ . Notice that 
the incident heat flux $F$ appears nonlinearly - directly as well as through $\eta({P})$ with ${P}= P(F)$.  Equation (\ref{Fick}) can be viewed as Fick's law or as an electric circuit equation where the Voltage difference $(\Delta T=T_1(P) - T_0)$ is equal to the product of the current $F$ and the resistance $\eta({P})$; the latter, of course,
reflects the properties of the layer plasma. More explicitly, $\eta({P})$ is an effective impedance (inverse diffusivity); the larger the $\eta({P)}$, the larger the temperature difference that can be maintained for the same heat inflow.

Determination of the function $\eta({P})$ will, necessarily, require rather involved analysis and computational studies of plasma dynamics. When YM08 was published, there were hardly any microscopic studies of plasma dynamics in the tokamak edge regions. So the layer analysis was carried out without any input from dynamical studies. However, we now have a considerable body of knowledge that we called MTB in the introduction, and we will  launch a two pronged attack on the layer dynamics combining $MTB$ and $T_hTB$.

Pursuing  $T_hTB$ (YM08), therefore, we proceed to construct an explicitly solvable but reasonable model to extract the necessary conditions on $\eta({P})$ that might produce a new state -a {\it Flow}-dominated transport barrier, for example.  

A simple parameterization of $\eta({P})$:
\begin{equation}
\eta({P}) = \eta_0 + \eta_1({P}) = \eta_0 + a {P},
\label{resistivity}
\end{equation}
where $a$ is a constant, turns out to capture the essence of the problem.  A positive $a$ (increasing the effective inverse diffusivity, and thus, decreasing the prevailing diffusivity) will prove to be the harbinger of a phase transition.

The base-line impedance $ \eta_0=\eta(0)$ characterizes the ambient state in the absence of {\it Flows}. 
Naturally this coefficient varies from system to system and is, in general, complicated. Fortunately, MTB will provide 
extremely useful estimates for $\eta_0$. 

For a broader scientific perspective, we recall that the ``diffusion'' in a tokamak, for example, 
is known to be driven by fluctuations yielding much higher (turbulent) heat-transfer rates as compared to the purely collisional ones. When applied to tokamaks, our fiducial (ground) state, then,  will be determined by  this very turbulent diffusive heat transport. This is where input from MTB will be crucial.

To develop the main features of our model, however, we do not need to know much about $\eta_0$; 
it is fully equivalent to specifying the reference inner boundary temperature and the reference layer temperature difference
\begin{equation}
T_D \equiv T_1(0)  = T_0 + \eta_0 F, \quad\quad \Delta T_D=T_D - T_0
\label{Fick2}
\end{equation}  
attained by the {\it Flow}-less ambient state (eventually to be identified with the L-mode).

We assume that the inter-relationships are defined by Fick's law; $F = D \Delta T
/\Delta x$ where $\Delta x$ is the layer thickness and 
$D~( = \Delta x/\eta_0)$ is the heat diffusion coefficient
(assuming a slab geometry and constant $D$, the heat flux
of the diffusion is $\bm{f}=-D\nabla T$ that is a constant vector).  The
entropy production associated with this diffusion process is
$\int\dot{S}_DdM = (T_0^{-1} -T_D^{-1})F$. 

In the next obvious step, using (\ref{Fick}), (\ref{resistivity}) and (\ref{Fick2}), 
we eliminate $\eta_0$ (in favor of $T_D$) to arrive at
\begin{equation}
T_1(P) =T_D + a P F.
\label{tempdiff}
\end{equation}
which will, naturally yield $T_1(P) >T_D $ for a positive $a$.


In order to extract the essential content of (\ref{tempdiff}), we must find an expression for the  power $P(F)$ available to generate the Flows. To do this, we must subtract the power wasted through the ubiquitous entropy
production in a diffusive process, $T_0 [T_0^{-1} -
T_D^{-1}]F $ (follows from Eq.(\ref{workdonetranstext}) with zero work done) from the maximum power ${P}_{\textrm{max}} = [1-T_0/T_1(P)] F$ available in an ideal Carnot process (\ref{entropy-emission}):
\begin{equation}
P = P_{\textrm{max}} - T_0 \left( \frac{1}{T_0}-\frac{1}{T_D}\right) F
= T_0 \left( \frac{1}{T_D}-\frac{1}{T_1(P)}\right) F .
\label{available_power}
\end{equation}
However, not all of this $P$ can appear as the Flow energy because of
inherent damping mechanisms (additional entropy production).
The coefficient $a$ multiplying $P$ in (\ref{resistivity}) was introduced just to reflect this fact; $1>a>0$ may be 
viewed as some sort of an efficiency factor, and scales the over-all influence of the Flow on the thermal
transport.

Equations  (\ref{tempdiff}) and (\ref{available_power}) are simultaneous in $T_1$
and $P$ (itself a function of $F$), we can solve them for either.  The quadratic, 
\begin{equation}
(T_1 -T_D) (1-aF^2\frac{T_0}{T_1T_D})=0,
\label{consistency1}
\end{equation}
yields
\begin{equation}
(T_1 -T_D) =0,
\label{sol1}
\end{equation}
the reference diffusive solution that pertains when ${F}=0$, and 
the flux driven solution
\begin{equation}
T_1  =aF^2T_0/T_D, \quad \frac{T_1}{T_D}  =\frac{aF^2T_0}{T_D^{2}}
\label{sol2}
\end{equation}
the very object of our investigation; the latter is capable of supporting a higher temperature
contrast ($T_1>T_D$) if 
 \begin{equation}
\frac{T_1}{T_D}  =\frac{aF^2T_0}{T_D^{2}}\equiv \frac{F^2}{F_{c}^2}>1, \quad F_{c} = \frac{T_D}{(aT_0)^{1/2}},
\label{sol3}
\end{equation}
where $F_c $ is the critical power that must be exceeded for bifurcation leading to the high gradient state.  Since ${T_D}$ is a function of $F_{c}$,  we could eliminate $T_D$(using (\ref{Fick2})) and express
\begin{equation}
F_{\textrm{\textrm{c}}}\equiv \frac{T_0}{\sqrt {T_0a} - \eta_0} .
\label{threshold2}
\end{equation}
in terms of the three fundamental parameters of the model $a$, $\eta_0$ and $T_0$.

Before we explore the implications of the preceding results, let us pretend that this model really pertains to tokamak experiments. When the layer dynamics is controlled fully by diffusion (turbulent plus collisional), then the system is an L-mode with a temperature $T_D$, and a temperature difference ${\Delta T}_L=T_D-T_0$ in the local region- there is no transport barrier. When the power influx crosses the threshold ($F>F_{c}$), the system transitions to an H-mode with a ${\Delta T}_H=T_1-T_0> {\Delta T}_L$ and there is an edge transport barrier. Thus the condition $F>F_{c}$ is the condition for an L-H transition and $F_{c}$ defines the threshold for an L-H transition arrived at by purely macroscopic "thermodynamical" reasoning. Naturally in $T_h$ TB, $F_{c}$ is undetermined.

\begin{figure}[H]
    \centering
    \includegraphics[scale=0.14]{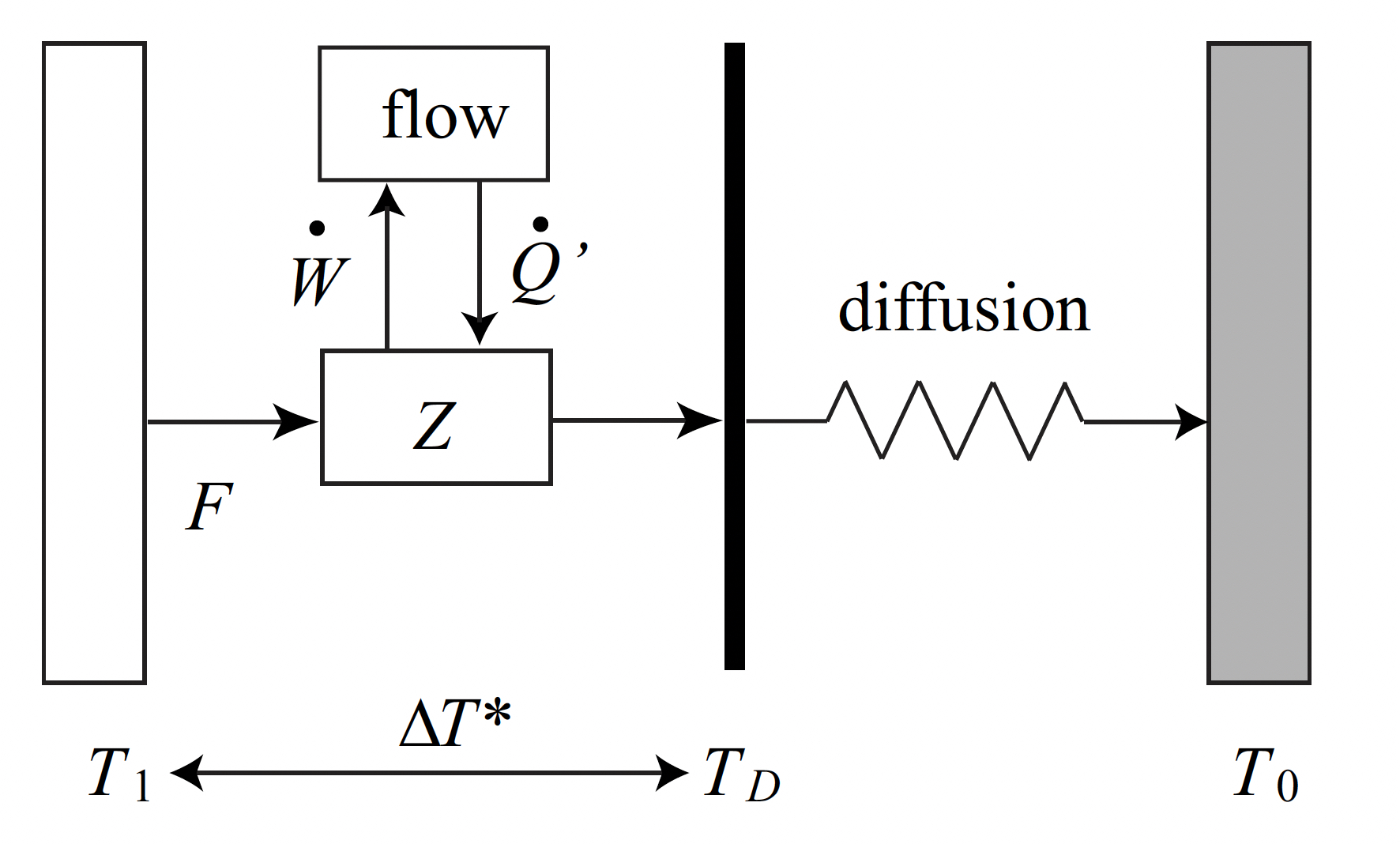}
    \caption{\label{fig:diagram} Equivalent diagram of a heat engine in a boundary layer. For ${\Delta T}> 0$,the heat engine can work to drive Flows (P is the power driving the flow which equals the dissipative power in steady state.). The flow produces
an additional “nonlinear impedance” $Z = \eta(P) P$ that sustains the temperature
contrast ${\Delta T}$ yielding a free energy to drive the flow itself. }
\end{figure}

First and foremost revelation from (\ref{threshold2}) is that for positive heat flow into the layer from the inner boundary ($F >0$), the threshold condition is meaningful only if the edge temperature $T_0$ is sufficiently high to satisfy
\begin{equation}
T_0 > \frac{\eta_0^2}{a} \equiv T_c
\label{condition_3}
\end{equation}
where $T_c$ is the critical temperature below which a transition is not possible. This is, indeed a striking prediction of this model - {\it that a very cold outer edge could
prevent the L-H transition even for arbitrary amounts of heat flux input into the layer}. 

The expression for $T_c$ seems totally commonsense- 
As $\eta_0$ becomes larger and the efficiency factor $a$ (of converting the available $P$ to do work) becomes smaller the resistance for non-diffusive processes to dominate becomes larger!

 We will come back to a longer discussion on the implications when we review these results in 
the light of the Microscopic-dynamical picture of the TB (MTB). 

It is essential to point out  that the bifurcation condition does not depend on the detailed shape of the function $\eta({P})$, as long as $\eta'({P})>0$.  For a general $\eta({P})$, bifurcation ((\ref{sol3})) will occur if
\begin{equation}
 \frac{\eta'(0) F^2 T_0}{T_D^2} > 1;
\label{condition_1'}
\end{equation}
replacing $a$ by $\eta'(0)$ in (\ref{threshold2}) and (\ref{condition_3}) will yield the minimum heat flux and the minimum temperature.

What happens if $\eta({P})$, though monotonically increasing function of P, saturates as $P \rightarrow \infty$. Let us examine 

$$\eta(P) = \eta_0 + \left[aP/(b+P)\right], $$ 
that is always finite; it is also monotonic and exhibits the properties: $\eta \rightarrow \eta_0$ for $P \rightarrow 0$ and $\eta \rightarrow \eta_0 + a$ as $P \rightarrow \infty$. It is straightforward to calculate that $F_min$ now turns out to be 
\begin{equation} \label{FcA2}
F_c = \frac{T_0}{\sqrt{T_0 a/b }-\eta_0}.
\end{equation}
which is effectively the same as (\ref{condition_3}).

\section{Stability of the bifurcated state}

Are the two states ($T_1=T_D$, the L mode and $T_1>T_D$, the H-mode, predicted in this layer model) stable? The stability analysis is done by examining the response of $T_1$ to a perturbation $\delta T$:  
\begin{eqnarray*}
\delta T \quad &\rightarrow & \quad \delta{P} = F \frac{T_0}{T_1^2}\delta T
\\
&\rightarrow & \quad \delta \eta = a F  \frac{T_0}{T_1^2}\delta T
\\
&\rightarrow & \quad \delta T_1 
= a F^2  \frac{T_0}{T_1^2}\delta T \equiv \alpha \delta T.
\end{eqnarray*}
Since $\delta T$ is amplified (diminished) for $\alpha >1$ ($\alpha <1$), the magnitude of $\alpha$ is the determinant of stability. Individually,

%
 \begin{itemize}
\item For the trivial state $T_1 = T_D =
T_0 + \eta_0F$,  below the bifurcation point, i.e., $aF^2T_0/T_D^2
\leq 1$ [see (\ref{sol2})],
\[
\alpha \equiv  a F^2  \frac{T_0}{T_1^2} \leq \frac{T_D^2}{T_1^2} = 1
\]
(equality holds at the bifurcation point), the trivial state is 
stable.  {\it For fluxes above the bifurcation threshold, however, the
trivial state becomes unstable}.  

\item On the other hand, the high $T_1= aF^2T_0/T_D  >T_D$ state, if it exists, is always stable,
because  
\[
\alpha \equiv  a F^2  \frac{T_0}{T_1^2} = \frac{T_D}{T_1} < 1.
\]
\end{itemize}
In this model, therefore, the very existence of H-mode guarantees its stability. 


\section{Multi-scale Flow-turbulence system}
\label{sec:multi-scale}

We are now all set to marry the Thermodynamic TB ($T_h$TB) model (constructed in YM08 and reproduced in the last few sections) with the vast  body of scientific knowledge (MTB) acquired via experimental and theoretical (advanced computing and semi-analytical solution of the microscopic plasma dynamics) investigations. But before that we must justify the workings of such a counterintuitive model like the $T_h$TB -  how can we create order (a TB with coherent motions) and disorder (maximum entropy production) in the same breath. 

Perhaps, the first step in the deeper enquiry into the workability of the model ``heat engine'' we are developing is the realization that the enhancement of temperature gradients arises because  the incoming energy, preferentially, generates Flows, minimizing the energy available for turbulence-driven diffusive heat transport. In the terminology of this work, this corresponds to an increase in effective impedance, $\eta(P)>\eta_0$. 

We remind the reader that we are in a territory very different (likely orthogonal) from that carved out by the “principle” of minimum entropy production, which is often used to explain the emergence of organized structures such as flows. Within that framework, high entropy production is typically associated with the destruction of coherence and a drift toward disorder.

Below we summarize the essentials of the content of this ``deeper enquiry'' (into workings of the heat engine) advanced in YM08-  The reader is requested to consult this reference and Appendix B for more details.

Functionally, the key element that distinguishes the formulation of current model lies in the boundary condition at the inner surface. Rather than prescribing a temperature $T_1$, we instead impose a fixed heat flux $F$ entering the layer. This flux supplies the energy that can ultimately be converted into coherent flow. Crucially, the system only accesses the high temperature-contrast state when $F$ exceeds a threshold; below this level, the “engine” does not operate.

Perhaps, at its very core, this enquiry illustrates the basic idea that the revealed dynamics is dictated by what we choose to be the ``Control'' thermodynamic variables. New dynamics does result in response to a change in the Control parameters (such  a change is mathematically described by a Legendre transformation). In the context of this paper, It is relevant to inform the reader that the generating function for the Legendre transformation connecting  flux-driven and force-driven systems has been shown to be the entropy production rate; the thermodynamic potential for which is a generalization of Onsager's dissipation function \cite{YK2014}.

Since the high temperature-contrast state is the final product of the heat engine, the factor measuring the temperature-contrast, $(1-T_0/T_1)$ sets the scales for the two processes --the Carnot efficiency for generating mechanical energy (Flows, order), and the production (emission) of entropy (disorder).

There is a neat and somewhat obvious solution to this concurrent order-disorder phenomena; the two opposing mechanisms could, indeed, act independently and simultaneously if the ``domains'' of their efficient operation were non-overlapping. Invoking Scale-separation will guarantee, precisely, the right conditions: 
\begin{itemize}
\item
the total entropy production is dominated by small scale perturbations with a large damping rate ($\propto L^{-2}$; $L$: eddy size) keeping the eddy amplitudes (sacrifice for the dissipation) to be very small,

\item In contrast, the Flow, a coherent macroscopic large-scale structure, is nonlinearly driven by an instability induced  by the entering heat flux $F$; its creation/characteristics are not affected by the short scale dissipation responsible for entropy production.
\end {itemize}


Thus the existence of a scale-hierarchy allows the system to transition to a state that can maintain order while maximizing disorder. This transition is an expression of self-organization of the ``heat engine''; the self-organization, most likely, taking place through the so-called ``dual-cascade'' process investigated in two-dimensional (2-D) turbulence. Using this approach, for instance, self-organization of``zonal flows'' in electrostatic turbulence of plasmas was predicted \cite{Hasegawa_Wakatani}

In appendix B, we discuss the dual cascade phenomenon via the canonical example of two-dimensional turbulence provided by the 2-D Navier-Stokes system~\cite{Hasegawa}. It is shown that in this system the dual cascade is facilitated by the existence of two different ideal constants of motion, the energy and the enstrophy. When fluctuations are excited (by an instability) at an intermediate range of wave numbers, the energy and the enstrophy move in opposite directions in the wave number space; the energy transfers, through the inverse cascade route, toward larger scales (gets ordered), while the enstrophy goes to the small-scale dissipation range (gets disordered). At the large-scale, a flow self-organizes, and the stretching effect suppresses turbulent transport.

Although in our system, we cannot isolate a second constant of motion, we do have a dynamic constraint that plays a role similar to enstrophy( Appendix B)- after all a constant of motion is a constraint on dynamics.

Strictly speaking, the only physical system for which the dual cascade induced self-organization is demonstrated is a 2-D fluid; in three-dimensions (3-D) the vortex-tube stretching effect violates the ideal conservation of the enstrophy. 

A 3-D tokamak plasma, however, has an advantage over the 3-D neutral fluid; the strong axial magnetic field imparts an effective 2-D behavior to the confined plasma so that simpler 2-D fluid like considerations could be relevant to H-mode layers and ITB's.

\section{ Understanding Transport Barriers- Theory and Experiment}

It will hardly be an exaggeration to declare that the controlled thermonuclear program via magnetic confinement was saved from obliteration when (in 1982) the ASDEX team found the H-mode~\cite{wagner82} - a high confinement configuration which was believed to be a self-organized state brought about by the suppression of the dominant source of turbulent transport; the remarkable self-organization was accompanied by a strong velocity shear flow in the plasma. In the following years , almost every tokamak was able to create such configurations that had an edge transport barrier- a region of high pressure gradient occupying the region $~1> R > .8$ of the discharge. Not all H-modes were exalted; they varied from confinement improvement (called the H factor) by a factor of over 2 to barely above 1. Though the experiment led theory but over the last 35 years an enormous amount of theoretical /computational work has been done ; in the latter, the properties of transport barriers were derived from first principles microphysics of plasmas. 

Much of the theoretical microphysics literature did not exist when YM08 investigated this thermodynamic oddity from a macroscopic perspective. Many general results of YM08 ($T_h$TB) can be now be examined, interpreted and better understood in terms of the detailed physics we have called MTB and by the same token, MTB can be examined, interpreted and better understood in terms of the very general macro-physics revealed in  ($T_h$TB) summarized here. This section is fully devoted to cross fertilizing MTB-$T_h$TB and building a more advanced/encompassing theory of transport barriers; this effort will also suggest what experiments/theory could be done to construct a more complete picture. In what follows, the "layer" of this paper will be used synonymously to a TB (internal or edge). 

The YM08 calculation has  three independent parameters that determine the layer response to the incoming heat flux $F$: $T_0$, the temperature of the heat bath, $\eta_0$ the purely diffusive impedance to the heat engine, and the number $a$ measuring the efficiency of the available power to do work (produce coherent plasma motions: Flows) - these parameters are some macroscopic labels that characterize a given plasma configuration. They are directly or indirectly related to experimentally observable quantities- they can also be computed  via first principle simulations. So the predictions of this model can both be qualitatively and quantitatively tested/corroborated by MTB. 

Let us now examine the fall out form the striking condition that for the L-H transition to occur (at all), the heat bath temperature $T_0> T_c$  where (\ref{condition_3}) 
$$ T_c = \frac{\eta_0^2}{a} $$
For $T_0<T_c$, no matter how hard you drive the system (by increasing $F$ ), the plasma will stay in the L mode. 
This is essentially an absolute criterion that MTB should be able to translate into requirements on measurable plasma parameters- This will advance the understanding of how to affect L-H transitions.

Much can be learnt, however, by examining, in more detail, the behavior of the critical power $F_c$ with the edge temperature $T_0$. It turns out that $F_c$ has a minimum($=F_{min}$) at a temperature we call ($T_{optimum}\equiv T_{opt}$). Therefore, in the vicinity of  $T_{opt}$, the $F_c -T_0$ curve is a parabola. Let us derive its defining parameters.

Differentiating (Equation(\ref{threshold2})) with $T_0$, 
\begin{equation}
\frac{\partial F_{c}}{\partial T_0}  = \frac{1}{2\sqrt {aT_0}} ( 1-\frac{\eta_0^2}{(\sqrt {aT_0}- \eta_0)^2},
\label{Fminprime}
\end{equation}
we find that the minimum occurs at 
\begin{equation}
T_0  = T_{opt}= \frac{4\eta_0^2}{a}=4T_c ,  \quad F_{min}=\frac{4\eta_0}{a}=\frac{T_{opt}}{\eta_0}
\label{Fminvalue}
\end{equation}
To display the behavior in the vicinity $T_{opt}$, one calculates 
\begin{equation}
(\frac{\partial^2 F_{c}}{\partial T_0^2})_{T_0=T_{opt}} = \frac{a}{8\eta_0^3}
\label{doubleprime}
\end{equation}
and finds 
\begin{equation}
F_{c}\approx F_{min}+ \frac{a}{8\eta_0^{3}}(T_0-T_{opt})^2=\frac{4\eta_0}{a}+\frac{a}{8\eta_0^{3}}(T_0-T_{opt})^2
\label{Fparabola}
\end{equation}
Both $F_{min}$ and the curvature ($=\eta_0^2/4$) of the ($F_c - T_0$) curve are experimentally measurable. We will soon test the model results against experiments.

Let us review the results:

\begin{itemize}
 \item As the bath temperature crosses the threshold $T_c$, but remains below $T_{opt}$=4 $T_c$ ~\cite{N2}, the minimum power needed for the transition decreases because $\partial F_{min}/\partial T_0$ is negative; in fact the rate of decrease is rather fast  because of the factor $ (\sqrt {aT_0}- \eta_0)^2 \propto(\sqrt T_0-\sqrt T_c)^2)$ in the denominator of (\ref{Fminprime}). We remind the reader that the lowest value $F^{min}$ is reached at $T_0=T_{opt}= 4 T_c$. On either side of $ T_{opt}=$, $F_{min}$ increases.
 
\item Lowering of $F_{min}$ has a direct effect on giving a higher enhancement factor 
\begin{equation}
h=\frac{T_1}{T_D}=\frac{F^2}{F_{min}^2} 
\label{enhancement1}
\end{equation}
for the same input power $F$. Or equivalently, at higher $T_0$ it takes less power $F$ for the same enhancement factor. in short, higher temperatures edges are fundamental boosters of confinement!
The truth of this statement is to be strongly tested by experiment as well as detailed dynamic simulations- the entire MTB edifice.

\item At $T_{opt}\equiv4 T_c$, the enhancement factor reaches its maximum value,
\begin{equation}
h=\frac{F^2}{(F_{min})^2}= \frac{a F^2}{4 T_{opt}},
\label{enhancement2}
\end{equation}
for a given set up - $T_{opt}$ and $a$ are characteristics of a given configuration and are related to quantities available in MTB.

\item The reader is surely aware that due to the stiff core transport(demonstrated strongly by gyro kinetic simulations) a higher $T_1$ leads to a commensurately high central temperature(for high fusion gain). And our very general macroscopic analysis prescribes that (for a give configuration) a higher $T_0$ may be the most effective route to higher $T_1$. 

\item That $T_0$ may be a major controller of overall confinement is a very robust result- Proven under minimum assumptions, the details of the dynamics of the layer (which set of instabilities are there, and how can they  be stabilized) do not change the basic phenomenon. However, it is precisely the MTB methodology that will tell us how to create conditions (magnetic geometry, heating mechanisms,  pellets, density control, ----) that will lead to a higher $T_0$- that facilitates the transition as well as accentuates the confinement gain.

\item The macroscopic $T_h$TB layer analysis is an effective predicting tool when the  trio of parameters ($T_0$, $\eta_0$, and $a$) are specified; these crucial physical parameters have to be found via MTB - In fact how we can engineer these parameters to be in the desired range has required painstaking work in which experiment-theory/computation have brilliantly combined. 

\end{itemize}

A very essential general message of $T_h$TB is that if the input power $F$ can effectively do mechanical work - generate coherent plasma motions (flows and currents) - then the  diffusive processes (turbulent and collisional) in the layer can be overpowered and states with enhanced  temperature inhomogeneity can emerge.  The exact pathway by which these {\it Flows} are created is not relevant to either the L-H transition threshold or the quality of the transport barrier

The association of H-modes with generation of plasma sheared flows is as old as the discovery of the H-mode itself.  $T_h$TB, though, predicts a causal connection, and gyrokinetic computations (MTB)show that the sheared flows damp the ITG-TEM (instability that would have been the major  source of turbulent diffusion)(see \cite {KLMHM}, for example). The experiment, the $Th$ TB and advanced dynamical theories all seem to agree and fortify one another.

Although the flows as causals (associates) of H-modes has wide spread recognition, the role of plasma currents as initiators, causals, or associates of H-mode is not so widely appreciated despite the predictions of both $T_h$TB and of detailed gyrokinetic (linear as well as nonlinear) calculations [see, for example, \cite {KLMHM}]. Because the exploitation of plasma currents could (immensely) extend the range of scenarios where transport barriers may be engineered, additional discussion is in order. 

In a multiple component plasma (typical magnetically confined system), the equivalent of the flow in a neutral fluid consists of  several flows, most recognizable being the bulk flow (ions and impurities) and the electrical current. Both of these are ordered plasma motions that  according to Ozawa {\it et al.}'s~\cite{Ozawa2001} demonstration could create high gradient states by suppressing instabilities. Thus, in addition to generating sheared flows, there exists a complementary pathway to creating TBs - induce a finite current; this mechanism will be particularly relevant in systems  in which creating sheared flows may be hard. In a very detailed paper, a subset of the authors, Kotschenreuther et al \cite {KLMHM}, have shown that a dynamical constraint  (derived from basic physics) could prevent plasma instabilities (in a conventional edge plasma) despite availability of free energy- The constraint is that  for gyrokinetic instabilities to exist, the total charge flux (current) must be zero. The implication is that any currents (coherent flow) will weaken/eliminate instabilities, diminishing turbulent diffusion, allowing TBs. Thus nonzero plasma current, just like the shear flow, reduces turbulent diffusion and sets the stage for the formation of a TB with boosted confinement. 

The most important aim of this paper was the demonstration of the grand equivalence of the general (but many) predictions of $T_h$TB and some of the conclusions reached by the highly impressive and epoch-making experimental and theoretical work on Transport Barriers. This agreement and mutual vindication give us encouragement  (because there exists a solid basis for the understanding of this remarkable state ) to advance the subject further, in particular, by  exploring new and yet unexplored configurations for maximizing confinement with all the engineering constraints.

The new enquiry could have, at least, two well defined objectives : 1) helping improve confinement in known configurations, 2) work out altogether new regimes where TBs could be engineered. We will speculate on both

\begin{itemize}
\item Since the edge temperature (let us identify $T_0=T_{sepratrix}\equiv T_{sep}$) is a prime booster of confinement (warranted by both approaches) we must find ways to increase it to its optimum value by whatever means available. The most direct route is via control of the edge particle source: the SOL density can reliably be increased with gas puffing and/or impurity seeding, and conversely it can be decreased by pumping or even implementation of a low recycling boundary.  This will naturally increase the SOL temperature since the same amount of power must be exhausted with fewer particles.  One of the indirect routes is illustrated in \cite {KLMHM} where by  lowering the edge density (lower $n_{sepratrix}\equiv n_{sep}$, higher density gradients) one could  increase $T_{sep}$ if the pressure were constant. However, $T_{sep}$ and $n_{sep}$ could, in general, be set through different knobs, and could be independently controlled. either of them (or together) could could aid the formation and boost the strength of TBs. The Macroscopic- Microscopic synthesis, attempted in this paper gives the experimentalist two alternative (or cooperative) knobs for for aiding the formation of and boosting the strength of TBs- a higher $T_{sep}$ and/or a lower $n_{sep}$. 

For the edge pressure constant scenario, in particular, lower $n_{sep}$ would imply higher $T_0$  that helps the TB in two ways- 1) It would trigger a TB if the initial temperature was too low (below $T_c$), and/or 2) it would allow a TB for a lower heating power. The density stabilization route can be made accessible by a variety of methods \cite {KLMHM}. There is ample experimental support for confinement increase with lower $n_{sep}$. 

\item There are several configurations in magnetic confinement devices (tokamks, stellarators--)  where it is difficult to create transport barriers (the ones where shear flow is inadequate, or where the magnetic shear is week (like in stellarators) for example)- the plasma is, then, not able to find the enhanced gradient state missing out on the concomitant boost in confinement. The general argument is that such configurations have instabilities that cannot be sufficiently tamed and hence the turbulence diffusion is too much to allow a high gradient state. Jacking up the edge temperature  to make $T_{sep}> T_c$ may be exactly what is needed to suppress diffusion and transition to a high confinement state. 

{\it From $T_h$TB perspective one could say that the lack of a TB is synonymous with the inability of the layer plasma to convert enough of the incoming power into coherent motions (the plasma flow and flow currents); the ambient diffusion, then, controls the profiles? If we could induce either momentum and/or current in the layer, the profile leveling effect of diffusion could be overcome. Higher $T_{sep}$ may be the best trigger.  } 

We get the same message from the detailed stability studies, guided by the zero current constraint, carried out and reported in detail in \cite {KLMHM}. Any departure from zero net current is stabilizing and reduces the instability transport; any induction of current can work towards TB formation.

This suggests, for instance, introducing an impurity pellet into the layer - impurities, due to their higher charge, are more effective in creating  current imbalance that can drive the system into the regime where the layer could support TBs.

Thus  based on doubly supported knowledge one could confidently engineer TBs in otherwise inhospitable regimes!
\end{itemize}

\section{Micro-TB Theory and Experiment}

In this section, we review empirical evidence as well as theoretical aspects of the `micro' transport barrier physics that may be consonant /dissonant with the present $T_h$TB theory.  This discussion will be split into two parts: (1) aspects consistent with the thermodynamics: e.g., high $T_0$ is beneficial.  (2) aspects consistent with the grand transport picture: any process that preferentially channels energy to 'flows'---i.e., increases a---as opposed to diffusive transport processes--i.e. $\eta_0$--is conducive to transport barriers.

\subsection{Thermodynamics}

There exists a large body of experimental evidence regarding the impact of SOL temperature and density on confinement~\cite{Strachan_1994,Kallenbach_1999,Majeski_2006,Maingi_2010,Maingi_2015,Osborne_2015,Stefanikova_2018,Frassinetti_2019,Frassinetti_2021,Battaglia_2020,Bourdelle_2023,lomanowski22,Boyle_2023,Jackson_2015}.  Since confinement in H-mode plasmas is closely linked to characteristics of the edge transport barrier, this connects directly to the transport barrier physics.  Typically the observations are reported in terms of SOL density, since that is the quantity that is most directly under external control.  However, some studies report on edge temperatures~\cite{lomanowski22}.  In either case, the observations are highly relevant in the context of the present theory, since SOL density and temperature are typically tightly anti-correlated.  Almost universally, the trend is consistent with the model theory developed and presented here: higher confinement is correlated with higher SOL temperature (i.e. lower SOL density) as long as we do not exceed $T_{opt}$. We review several observations.  

\begin{figure}[H]
    \centering
    \includegraphics[scale=0.25]{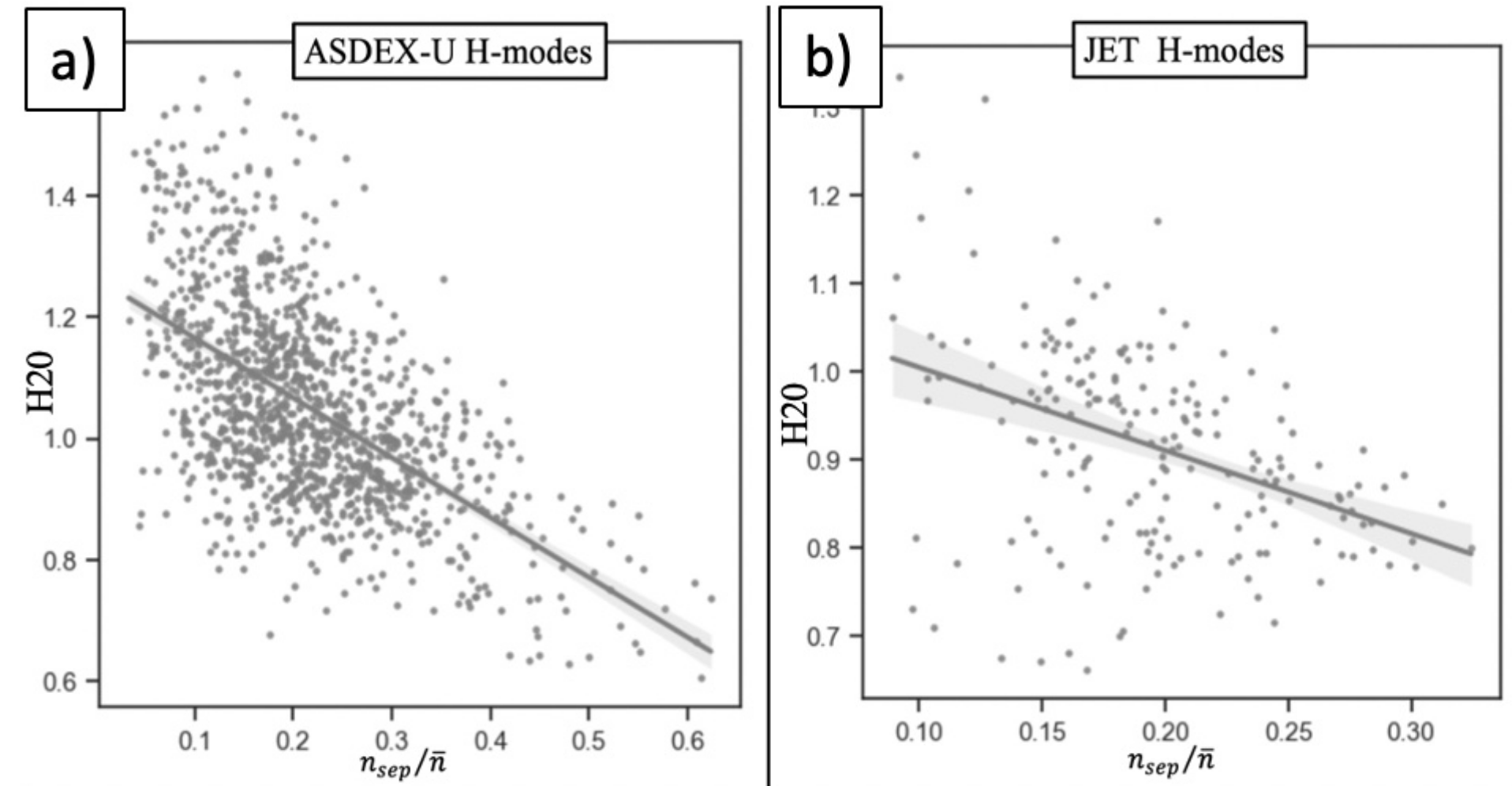}
    \caption{\label{fig:itpa} ITPA from Ref.~\cite{Kotschenreuther_2024}.   }
\end{figure}

\begin{itemize}

\item The NSTX experimental program is pursuing low recycling edge scenarios, wherein the boundary is conditioned with lithium, thereby absorbing incident hydrogenic particles.  This decreases the edge density and increases the edge temperature resulting in the highest performing NSTX discharges observed to date: this class of configurations is a major focus of the upgrade (NSTX-U).  

\item When JET implemented an ITER-like wall (JET-ILW), increased gas puffing was required to mitigate sputtering from the tungsten divertor.  JET performance declined substantially and several campaigns representing substantial operational innovation were required to recover performance approximating earlier C operation. All W mitigation strategies had to deal with the effects of increased in SOL density and corresponding decrease in SOL temperature. One of the main requirements for recovering performance was to maximize heating power, which, among many other things, would result in increased SOL temperatures.  A direct study of SOL temperature determined a strong dependence of confinement on SOL temperature near the divertor target (see Fig. 6 in~\cite{lomanowski22}).  In general, many characteristics of the JET-ILW confinement changes have been traced back to effects of SOL temperature and/or density~\cite{Stefanikova_2018,Frassinetti_2019,Frassinetti_2021,JETILW,Hatch_2019,Chapman_2022}

\item A strong connection between confinement and SOL parameters was also reported in WEST~\cite{Bourdelle_2023}.  Although these WEST discharges were in L-mode, it may still be indicative of the outer edge plasma temperature achieved in L-mode, which would also strongly impact confinement (note that there also exists a strong connection between total confinement and outer-edge T in L-mode discharges~\cite{pablo_24}.  

\item Perhaps the clearest exposition of the empirical connection between confinement and edge parameters is reported in~\cite{Kotschenreuther_2024}, which uses the latest ITPA H-mode confinement database~\cite{verdoolaege_21} to investigate these connections.  A representative figure is reproduced here in Fig.~\ref{fig:itpa}.  Notably, Ref.~\cite{Kotschenreuther_2024} also demonstrates that the {\it observed confinement trends can largely be attributed to transport properties in the barrier as opposed to exclusively MHD limits.}  
\end{itemize}

\subsection{Transport}

In this subsection, we review connections between detailed (micro) empirical and theoretical aspects of transport barriers and the main 'transport' thesis of the present theory, namely that the propensity for the system to generate Flows or other non-diffusive processes fosters the formation of transport barriers.  This is encapsulated in the parameter $a$ in Eq.~\ref{resistivity}.  Notably, this theory doesn't distinguish between fluid flows and currents, which substantially broadens the connections with known `micro' transport barrier understanding.  

The most obvious and well-known connection is the role of zonal flows--i.e. poloidal flows generated dynamically by the turbulence--in triggering the LH transition.  Zonal flows likely represent a major contribution to $a$ at the point of the LH-transition. This mechanism is the predominant feature of most TB theories and so we won't dwell on it beyond noting the following citations~\cite{terry_2000,diamond_2005}.  

At some point beyond the LH transition, the turbulence is largely suppressed and characteristics of the transport barrier are determined by other mechanisms.  These mechanisms can also be conceptualized as a propensity for the system to produce ordered flows; these could originate even via instabilities.  Returning again, to Ref.~\cite{Kotschenreuther_2024}, the density gradient in the barrier is perhaps the most important factor determining the quality of the TB in its advanced state as manifest by the ultimate confinement attainable via the TB.  In the context of the ThTB theory, many of the phenomena In this more advanced TB state can be attributed to the role of the density gradient in generating macroscopic flows (i.e. to enhance $a$).  Once the transition occurs, an $E_r$ well develops, which continues to suppress any residual turbulence in the barrier via $E \times B$ shear.  This $E_r$ well is described by neoclassical force balance, which contains the density gradient (consequently, the $E \times B$ shear has a factor of density gradient squared).  Also, the plasma current in the transport barrier is dominated by the bootstrap current, which can be large enough to substantially reduce the local magnetic shear.  This reduced magnetic shear is crucial for enabling access to second stability for kinetic ballooning modes (KBM), which allow for steep pressure gradient and higher confinement.  Likewise, a sufficiently large bootstrap current is necessary to provide access to the 'nose` of the peeling-ballooning stability boundary, which is crucial for maximizing pedestal pressure and confinement.  As with neoclassical $E_r$  the density gradient enters directly into the neoclassical bootstrap current and is a major factor in driving a large bootstrap current in the pedestal.  

Finally, perhaps the largest factor determining the quality of a TB from a `micro' TB perspective, is the capacity of density gradients to suppress the conventional turbulent transport mechanisms, ITG, ETG, and TEM as described in Ref.~\cite{KLMHM}.  In that work, turbulence reduction is enabled by two conspiring effects.  First, a favorable magnetic geometry must be realized in order to minimize the coupling of trapped electrons to the dynamics.  This effectively diminishes the capacity of the instabilities to drive particle transport.  The second effect is strong density gradients, which are thermodynamically predisposed to driving particle fluxes.  The only way to resolve the tension between the weak particle transport enforced by the system and the strong thermodynamic drive is for the instabilities themselves to become rather impotent---i.e., to be limited to small growth rates.  The following points will compare and contrast the current ThTB theory with the `micro' TB picture from Ref~\cite{KLMHM}:
\begin{enumerate}

\item{According to~\cite{KLMHM}, the key parameter in this system is fraction of the pressure gradient supplied by the density gradient: $F_p = \frac{a/L_n}{a/L_n + a/L_T}$ (note the close connection between $F_p$ and the more standard parameter $\eta = \frac{a/L_T}{a/L_n}$).  One common characteristic of the proper magnetic geometry is low magnetic shear $\hat{s}$, which can be driven by a strong density gradient via the bootstrap current, consistent with the ThTB picture of the system driving macroscopic currents and flows.}
\item{The picture in~\cite{KLMHM} can be viewed as yet another effect augmenting the `aP' term in the transport equation.  The steep density gradients emphasized in~\cite{KLMHM} accomplish exactly the physics encapsulated in Eq.~\ref{resistivity}: they enable transport to decrease as the power is increased and gradients steepen.}
\item{Both ThTB and \cite{KLMHM} point to separatrix temperature and density as key parameters.  One effect is directly related to the transport considerations of $F_p$ stabilization: If the TB is pressure gradient limited (e.g., by MHD stability), high Tsep and low nsep allow the system to reach the pressure gradient limit while retaining a low value of $F_p$ and hence low transport.  This aspect is entirely consistent with the ThTB picture encapsulated in Eq.~\ref{resistivity} as described above.  However, the ThTB picture also suggests that Tsep plays a role beyond the transport dynamics.  That is, ThTB suggests that for {\it fixed transport characteristics}, the confinement sensitively dependent on $T_{sep}$ as described above.  In other words, there is a thermodynamic consideration beyond the pure transport effects described in~\cite{KLMHM}.  One resulting prediction is a non-monotonic dependence of LH threshold power on $T_{sep}$, which is investigated in the next section in comparison with experimental data from Alcator C-Mod. }
\end{enumerate}

\section{Connecting with Experimental Data}
\label{experiment}

\begin{figure}[H]
    \centering
    \includegraphics[scale=0.65]{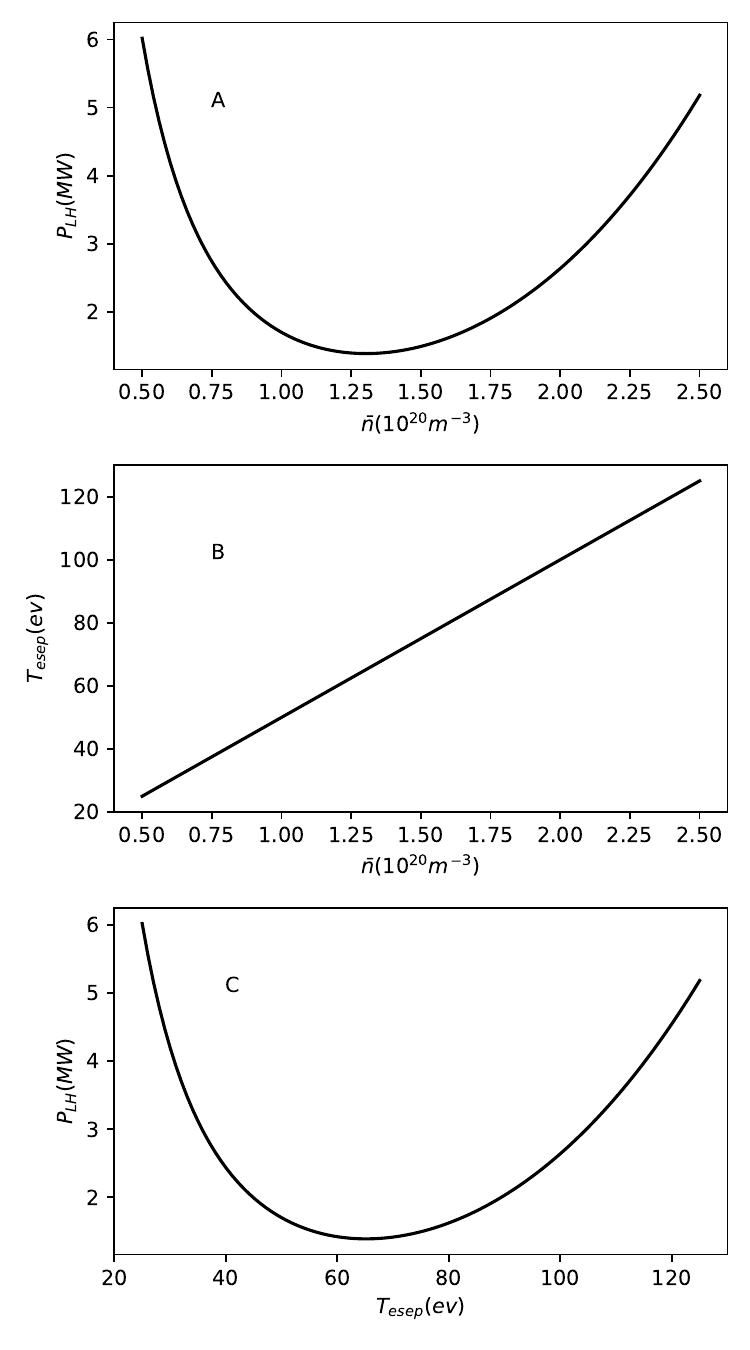}
    \caption{\label{fig:LH_quantities} This plot shows parameterizations of data from Alcator C-Mod published in Ref.~\cite{Cmod}.  The LH threshold power is shown in Fig. A as a function of line-averaged density (compare with Fig. 4 from Ref.~\cite{Cmod}).  The parameterization, $P_{LH} = 1.5/\bar{n}^2  + 0.2\bar{n}^{3.5}$,  is taken directly from Ref.~\cite{Cmod}.  Fig. B shows the relation between the separatrix temperature and the line-averaged density (compare with Fig. 7 D from Ref.~\cite{Cmod}), and Fig. C converts the data into the relation between the threshold power and separatrix temperature, making it directly relevant for the theory presented here.   }
\end{figure}

\begin{figure}[H]
    \centering
    \includegraphics[scale=0.65]{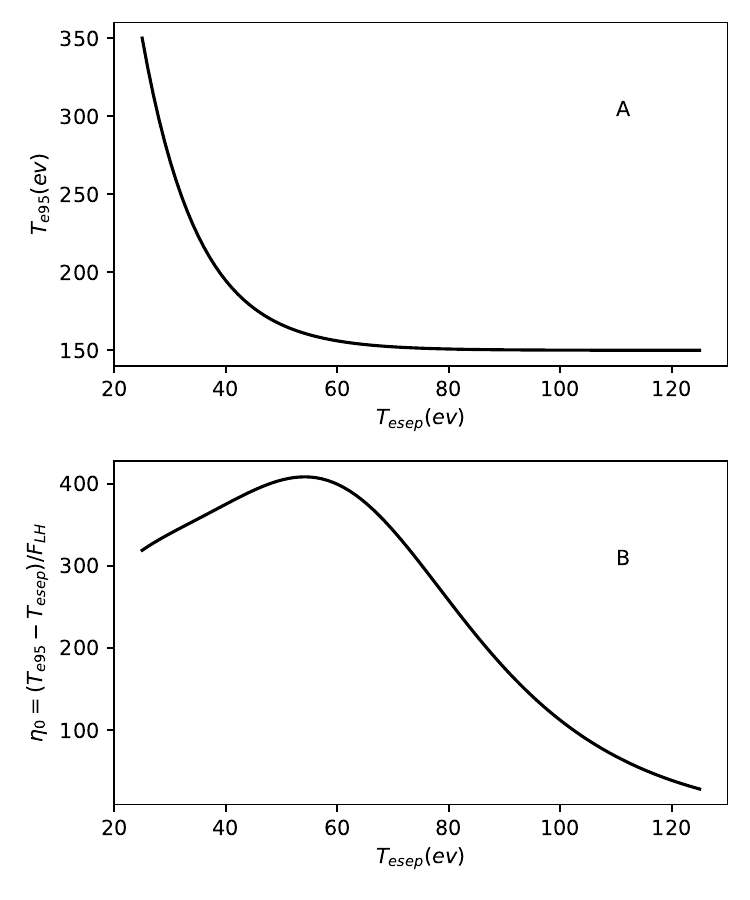}
    \caption{\label{fig:eta0} A. shows the parameterization of the electron electron temperature at $\Psi_N = 0.95$ at the LH-transition (parameterized to match Fig. 7 A in Ref.~\cite{Cmod}).  Along with the separatrix temperature and threshold power, this allows for the calculation of the threshold resistivity, $\eta_0$ (recall Eq.~\ref{resistivity}), shown in B.   }
\end{figure}

Finally, we will subject the present theory to the crucial test from experimental data; the latter is from Alcator C-Mod published in Ref.~\cite{Cmod}.  Ref.~\cite{Cmod} documents the dependence of the LH transition power on electron density, demonstrating a `U-shaped' dependence distinct from the standard monotonic dependence often cited in scaling laws~\cite{LHscaling}.  The data also includes separatrix electron temperature and electron temperature at $\psi_N=0.95$ (i.e., a proxy for $T_D$), allowing for direct comparisons with certain aspects of the present theory.  Fig.~\ref{fig:LH_quantities} shows parameterizations of data from Ref.~\cite{Cmod}: LH transition power as a function of $\bar{n_e}$ (Fig.~\ref{fig:LH_quantities} A), and the relation between $\bar{n_e}$ and $T_{e,sep}$ (Fig.~\ref{fig:LH_quantities} B).  Using this data, the relation between LH transition power and $T_{e,sep}$ (i.e., $T_0$) can be inferred as shown in Fig.~\ref{fig:LH_quantities} C.  Fig.~\ref{fig:eta0} shows $T_{e,95}$, the electron temperature at $\psi_N=0.95$, which then allows for the calculation of $\eta_0 = (T_{e,95} - T_{e,sep})/F_{LH}$.     

The parameterization of the threshold power $P_{LH} = 1.5/\bar{n}_e^2  + 0.2\bar{n}_e^{3.5}$ is prescribed in Ref.~\cite{Cmod}.  The other quantities have been parameterized to visually match the figures in Ref.~\cite{Cmod} as follows: $T_{e,sep} = 50\bar{n}_e$, and $T_{e,95} = 200 e^{-(\bar{n}_e-0.5)/0.2}+150$.  The heat flux $F_{LH} = P_{LH}/A$ is derived from the threshold power using a rough estimate of the boundary flux surface area $A = 2\pi a \times 2 \pi R = 5.9$ with $a=0.22 m$ and $R = 0.68 m$.   

\begin{figure}[H]
    \centering
    \includegraphics[scale=0.65]{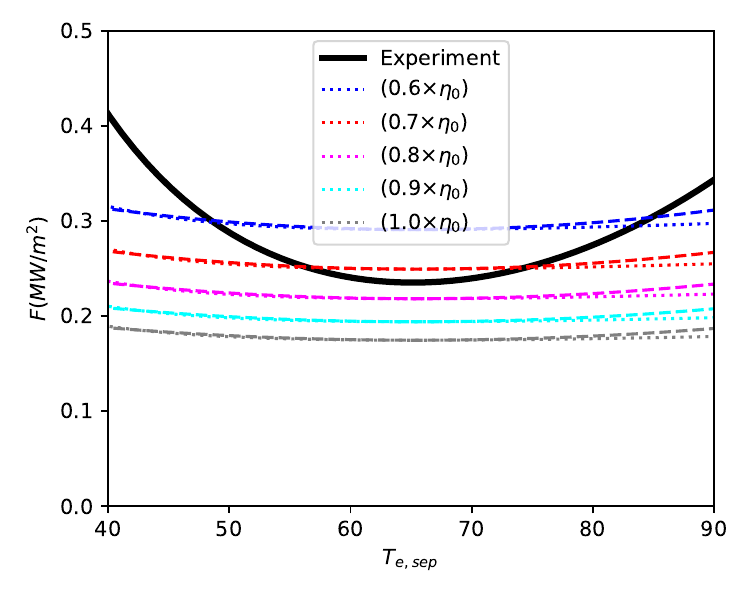}
    \caption{\label{fig:F_exp_vs_theory}.  A comparison of the experimental dependence (black) of heat flux at the LH transition and the theoretical predictions from Eq.~\ref{threshold2} (solid colors) and Eq.~\ref{Fparabola} (dashed color).  The theoretical predictions are shown for a range of values of $\eta_0$ justified by the intrinsic uncertainties in the experimental gradients.  The theoretical prediction of $F_{min}$ matches the experimental value for a $\approx15\%$ reduction in $\eta_0$.  However, the curvature is substantially under-predicted by the theory. }
\end{figure} 


This data directly confirms one major qualitative prediction of the theory: a minimum in the threshold power at a temperature substantially above $T_c$, as demonstrated in Fig.~\ref{fig:LH_quantities} C.  One can go further and compare the minimum value of the threshold power, by using Eq.~\ref{threshold2} in conjunction with the predicted value of the optimum temperature, Eq.~\ref{Fminvalue}.  We take the experimental value of $T_{opt} \approx 65 eV$, and the corresponding value of $\eta_0$ at this temperature and use Eq.~\ref{Fminvalue} to derive the corresponding value of $a$.  These values of $a(T_{opt})$ and $\eta_0(T_{opt})$ are then used in Eq~\ref{threshold2} to plot the behavior of $F_c$ in the vicinity of $T_{opt}$ as shown in Fig.~\ref{fig:F_exp_vs_theory} (the expansion derived in Eq.~\ref{Fparabola} is also shown for reference).  In light of the well-know uncertainties in experimental gradients, we plot $F_c$ for a several values of $\eta_0$ ranging from $60\%$ to $100\%$ of the nominal experimental value.  The prediction for $F_{min}$ is quite close to the experimental value, required a reduction of $\approx 15\%$ to match the data.  However, the curvature of $F_c$ is substantially under-predicted by the theory.

This comparison should be viewed as a preliminary investigation.  As discussed in the previous section, the parameters $a$ and $\eta_0$ are also sensitively dependent on $T_{e,sep}$ and consequently a more quantitative comparison would require us to take those dependencies into account as well.  Nonetheless, the comparison described here does validate two aspects of the theory: (1) a minimum in threshold power, and (2) a reasonably accurate prediction of the threshold power at this minimum. 



\section{ Summary/Conclusions}

Trying to construct a theory of the creation and sustenance of the high gradient states (HGS) is a daunting challenge - it is, however, essential to ``solve'' this problem because of the universal presence of such states- in physics as well as in biology. In this paper we have initiated an attack on this problem in the limited context of a  transport barrier in magnetically confined plasmas. We have proposed and solved a minimal macroscopic thermodynamic model ($T_h$TB) that deals with the problem of an idealized layer whose one end is maintained at a temperature $T_0$ and a heat flux F enters from the other end whose temperature $T_1$ is to be determined by invoking the principle of maximum entropy. The state is considered high gradient if 
$T_1- T_0 >T_D - T_0$,  where $T_D$ will be the temperature if the layer is diffusion dominated.

Interestingly enough, with very little input we can not only predict the conditions for the existence of an HGS (both $F$ and $T_0$ have to  be above some critical values $F_c$ and $T_c$) but give a more detailed behavior that could be compared with experiment as well as with detailed gyro kinetic simulations of the transport barriers (MTB).  

The reader is encouraged to read the text for details but we give here a two of the salient findings:

\begin{itemize}

\item The deeper physics for the existence of such states lies in, perhaps the ``obvious'' verity (in hindsight), that when coherent processes like the fluid flows and currents are preferentially generated by the incident power, HGS are favored as compared to the diffusion dominated normal states. In the ${T_h}$TB 
the mechanism of the generation of Flows is not relevant although that will be one of the questions  that MTB can and must answer. $T_h TB$, however, tells that when $F$ exceeds the threshold, the high gradient state is stable while the diffusive state is not.

\item The maximum confinement enhancement $h=F/F_c$ occurs when the edge temperature temperature is $T_{opt}=4T_c$ - that is where the threshold power  $F_c$ has a minimum. This parabolic (U shaped) behavior, though definitively verified by experiments, is not a part of the normal lore. 

\end{itemize}

Finally, all qualitative predictions of $T_h$TB are fully supported by MTB and experiment.  The prediction of a non-monotonic dependence of $F_c$ is one clear example of the predictive power of this theory, which heretofore has not been discovered by MTB.  This work sets the stage for more comprehensive studies of transport barriers and the LH transition.  Fruitful next steps could include (1) accounting for separate thermodynamic channels for electrons and ions, (2) including density dynamics, and (3) investigating the temperature (and density) dependence of $\eta_0$ and $a$.  

%
%
%
%
%
%
%

\acknowledgments 
The work of ZY was supported by Grant-in-Aid for Scientific Research
from the Japanese Ministry of Education, Science and Culture
No. 14102033 and No. 19340170.  SMM and DRH were supported by the U.S. Department of
Energy Grant DE-FG02-04ER54742.

S. M. Mahajan and D. R. Hatch report a financial and leadership relationship with ExoFusion, a fusion-energy company. This relationship did not influence the design, execution, analysis, or interpretation of this research. The terms of this relationship have been reviewed and approved by The University of Texas at Austin in accordance with its conflict-of-interest policies.

\section{Appendix A.  Thermodynamic Preliminaries }
 
 We will use the conventional notation: the changes in state variables are perfect differentials dU (internal energy) and dS (entropy ) while $\delta Q$ ($\delta W$) will denote heat absorbed (work done). Combining the first ($dU = \delta Q - \delta W$) and  the second law ($\delta Q = T(dS - \delta S_i)$ where  $\delta S_i (\geq 0)$ represents internal entropy production), one may write the (differential) work done as
\begin{eqnarray}
\delta W
&=& \delta Q + T_{ref} dS - (dU - T_{ref} dS)
\nonumber 
\\ 
&=& \left( 1-\frac{T_{ref}}{T}\right)\delta Q - T_{ref} \delta S_i -(dU - T_{ref}dS).
\label{Thlaws_1}
\end{eqnarray}
where a positive constant $T_{ref}$, measuring a reference temperature, has been introduced.

The first term on the right-hand side of (\ref{Thlaws_1}) gives the maximum work achieved in a reversible process (Carnot's theorem).  The second term ($-T_{ref} \delta S_i \leq 0$), proportional to the internal entropy
production, diminishes $\delta W$ in an irreversible process.The third term ($dU - T_{ref}dS$), consisting of exact forms, do not contribute to the integral over any closed cycle.

For an open fluid (plasma) system, one could define the thermodynamic
variables ($\delta W$, $\delta Q$, $U$, $S$ etc.) for each mass
element; these will be called ``specific energy'', ``specific
entropy'', and so on.  Their variations ($dX$ or $\delta Y$) are calculated
along the streamline.  Writing the time derivative of an exact
(non-exact) variable as $dX/dt$ ($\dot{Y}$), the rate of work done is
found from (\ref{Thlaws_1}):
\begin{equation}
\dot{W} 
= \left( 1-\frac{T_{\textrm{ref }}}{T}\right) \dot{Q} - T_{\textrm{ref }} \dot{S}_i 
- \frac{d}{dt}(U - T_{\textrm{ref }} S ).
\label{workdone1}
\end{equation}

Integrating (\ref{workdone1}) over a domain $\Omega$ (fixed) will give a macroscopic energy balance relation. Although a little involved, we will summarize the necessary steps. 

Denoting by $dM$ the mass element, the total amount of some state variable $X$ (evaluated
for a unit mass) is given by $\overline{X} = \int XdM$.  Note that this representation uses the Lagrangian frame ($dM$ moves with the
fluid).   

The time derivative of some integrated state variable $\overline{X}= \int XdM$, 
 where the mass element $dM=\rho d^3x$ ($\rho$ is the density and $d^3x$ is the volume element),
 may be manipulated as (using the mass conservation law $\partial \rho/ \partial t + \nabla\cdot(\bm{v}
\rho) = 0 $ ($\bm{v}$ is the flow velocity),
\begin{eqnarray}
\frac{d}{dt}\overline{X} &=& 
\int_\Omega \frac{\partial}{\partial t} (X \rho)d^3 x
\nonumber
\\
&=&
\int_\Omega \left[ \frac{\partial}{\partial t} (X \rho) +
\nabla\cdot(\bm{v} X \rho) \right] d^3 x - \int_{\Omega} \nabla\cdot(\bm{v} X \rho) d^3 x 
\nonumber
\nonumber
\\
&=&
\int_\Omega \left( \frac{\partial}{\partial t} X +
\bm{v}\cdot\nabla X \right)\rho d^3 x - \int_{\partial\Omega} (\bm{n}\cdot\bm{v}) \rho X d^2 x
\nonumber
\\
 &=& \int_\Omega  \left( \frac{d}{dt} X \right)dM 
- \int_{\partial\Omega} (\bm{n}\cdot\bm{v}) \rho X d^2 x 
\label{Eulerian_representation}
\end{eqnarray}
where $\bm{n}$ is the unit normal vector, directed outward, on the
boundary $\partial\Omega$, and $dX/dt = \partial X/ \partial t +
\bm{v}\cdot\nabla X$ is the convective (Lagrangian) derivative.  If we
assume that the mass flow is confined in the domain,
$(\bm{n}\cdot\bm{v})\rho X$ must vanish on the boundary.  In what
follows, we omit the mass flow through the boundary.

In a ``quasi-stationary state'' (could be far from thermal
equilibrium), a sufficiently long-term average of a state variable
must be constant.  Hence, we may assume that the volume integral of
the state variables ($U$ and $S$) are constant.  Integrating
(\ref{Thlaws_1}) over all fluid elements, then, yields
\begin{equation}
\int \dot{W} dM 
= \int \left( 1-\frac{T_{\textrm{ref }}}{T}\right) \dot{Q} dM
-T_{\textrm{ref }} \int \dot{S}_i dM .
\label{workdonerateof}
\end{equation}

Generally, the variations of non-exact variables may take finite
values even in a quasi-stationary state.  Indeed, (\ref{workdone1})
gives the estimate of the long-term average work (power) of a
quasi-stationary thermodynamic engine.

\subsection{Quasi-static layer system}

We are now ready to "derive" the  thermodynamics of an idealized plasma ``layer''
bounded from the inside by an internal core plasma, and from the
outside by a cold heat bath. What is specified is  the total heat
flux $F_1$  through the inner boundary $\Gamma_1$  and the temperature $T_0$ of the heat bath
at the outer boundary $\Gamma_0$- what is to be determined is the inner-boundary temperature $T_1$ 
whose value measures the layer temperature gradients.

The outer-boundary heat flux $F_0$ must balance  $F_1$ in a quasi-steady state (then, we write $F_1=F_0 =F$).

Neglecting mass flow across both boundaries, the boundary terms in (\ref{Eulerian_representation}) go to zero.  We
assume that $\dot{W}$ works only internally to drive  flows-currents in
$\Omega$ (the energy transformation between the thermal energy and the
mechanical/electrical energy of collective motion may be represented by $\dot{W}$
or its dual $\dot{Q}=-\dot{W}$).  The entropy production $\dot{S}_i$
is, by definition, internal in the domain.  However, the layer may exchange 
heat $\dot{Q}$ with the exterior.  

In terms of the heat flow vector $\bm{f}$ ($\dot{Q}\rho = -\nabla\cdot\bm{f}$),
after some manipulation, the first term of Eq.(\ref{workdonerateof}) may be written as ($dM=\rho d^3x$)
\begin{eqnarray}
& & \int_\Omega \left( \frac{1}{T_{\textrm{ref }}}-\frac{1}{T}\right) \dot{Q} \rho d^3x
\nonumber
\\
& & ~= - \int_{\partial\Omega} \left( \frac{1}{T_{\textrm{ref }}}-\frac{1}{T}\right) 
\bm{n}\cdot\bm{f} d^2x
- \int_\Omega \bm{f}\cdot\nabla\left(\frac{1}{T}\right)d^3x
\nonumber
\\
& & ~= \left[ \left( \frac{1}{T_{\textrm{ref }}}- \frac{1}{T_1} \right)F_1
- \left( \frac{1}{T_{\textrm{ref }}}- \frac{1}{T_0} \right)F_0 \right]
- \int_\Omega \dot{S}_DdM,
\label{entropy-emission}
\end{eqnarray}
where we have denoted $[\bm{f}\cdot\nabla(1/T)] = \dot{S}_D\rho$
($T$ and $\bm{n}\cdot\bm{f}$ are assumed
to be constant on both boundaries).
The first term of (\ref{entropy-emission}) represents the ``entropy emission rate''
through the boundaries.
The second term in (\ref{entropy-emission}) is the ``entropy
production rate'' due to the (irreversible) energy Flows $\bm{f}$. 

Hereafter, we set the reference temperature $T_{\textrm{ref} }=T_0$ (the
heat bath temperature).
Using (\ref{entropy-emission}) transforms (\ref{workdonerateof}) to
\begin{equation}
\int \dot{W} dM 
= \left( 1- \frac{T_0}{T_1} \right)F_1
-T_0 \int \left( \dot{S}_D+\dot{S}_i \right)dM .
\label{workdonetrans}
\end{equation}

If the heat were to transport only by diffusion in a stationary medium
(viz., $\dot{W}=0$), the entropy production (and, thus, the entropy
emission) is minimized~\cite{Berbera} for the ``harmonic heat flow''
($\nabla\cdot\bm{f}=0$), i.e., (\ref{workdonetrans}) holds with
$\dot{W}= 0$ and $\dot{S}_i=0$.

In a general quasi-steady state, the mechanical plus electrical energy (Flow energy; in the rest of the paper, the work done, and the excitation-dissipation Flows will be used exchangeably) must
saturate, and thus, $\int \dot{W}dM =0$ (by the first law of thermodynamics,
$\int \dot{W}dM = F_1-F_0$, and, in a quasi-steady state, $F_1=F_0=F$).
However, local $\dot{W}$ may remain
nonzero.  Excitation ($\dot{W}>0$) and dissipation ($\dot{W}<0$) of
energy in flows-currents may occur at different space-time locations, and possibly, at
different scales.


In the macroscopic energy balance equation (\ref{workdonetrans}), the influence of the Flows
(mechanical/electrical energy) on heat transport may be accounted by the entropy production term
$\int(\dot{S}_D+\dot{S}_i)dM$. In an initial transient phase, $F_0$ may be smaller than
$F_1$ and there is power available to generate the Flows ($\int \dot{W}dM = F_1-F_0$).  Then, entropy production will increase to balance the ``heat-engine drive'' as a quasi-steady state is approached.

The ``maximum entropy production'' ansatz will demand that
internal entropy production term must acquire the largest accessible
value.  Although in a transient phase, the entropy production rate
(energy dissipation rate) may assume any arbitrary value, in a
quasi-stationary state, it is bound by the total entropy emission
(that is proportional to the first term on the right-hand side of
(\ref{workdonetrans})); the factor $(1-T_0/T_1 )$ insures that the
entropy emission increases as the difference between $T_1$ and $T_0$
increases. {\it  Maximum entropy production is, therefore, fundamentally
tied to maximum temperature inhomogeneity.}  

Since the maximum work done  by the heat engine(synonymous with changes in the plasma
energy) also scales with factor $(1-T_0/T_1 )$,  the maximum entropy production and the maximum work done
are controlled by the same temperature difference factor, it follows
that entropy production will be maximized, if in the layer, a large
temperature inhomogeneity is excited/maintained/accompanied by large
plasma Flows.  Such a quasi-state, if found to be stable, will be
surely far from thermal equilibrium, and will need to be sustained by
an external input, for example, the heat flux entering the layer from
the core plasma.

\section{Appendix B.  Mechanisms for scale hierarchy }
 
 The canonical example of two-dimensional turbulence, however, is
provided by the 2-D Navier-Stokes system~\cite{Hasegawa} where the dual
cascade is facilitated by the existence of two different ideal
constants of motion, the energy and the enstrophy. When fluctuations
are excited (by an instability) at an intermediate range of wave
numbers, the energy and the enstrophy move in opposite directions in
the wave number: the energy transfers, through the inverse cascade
route, toward larger scales (gets ordered), while the enstrophy goes
to the small-scale dissipation range (gets disordered).  At the
large-scale, a flow self-organizes, and the stretching effect
suppresses turbulent transport. It should be noted that these
processes have been shown to function only in a 2-D fluid;
in three-dimensions (3-D) the vortex-tube stretching effect violates the
ideal conservation of the enstrophy. A 3-D tokamak plasma, however,
has an advantage over the 3-D neutral fluid; the strong axial magnetic
field imparts an effective 2-D behavior to the confined
plasma so that simpler 2-D fluid like considerations could be relevant
to H-mode layers and ITB's.

In the standard approach for exploring the large-scale structure of
the flow, one sets up a variational principle --a constrained
minimization of enstrophy while keeping the energy constant (as well
as the total angular momentum)~\cite{Hasegawa_Wakatani,Hasegawa}. The
minimum enstrophy principles, naturally, implies ``minimum entropy
production'' because the dissipation is proportional to the enstrophy.

In our model, however, the conventional variation principle is
essentially stood on its head; the relevant new principle will
constitute a dual~\cite{dual-VP} or an antitheses of the old one; one
must maximize the energy for some reference value of the enstrophy.
As mentioned above, the boundary condition --specifying the heat flux--
is the key to our departure from the standard picture and, hence, the
cause for the new variational principle.  In an open system where the
entering heat flux $F$ is given, the entropy production rate is
bounded:
\begin{eqnarray*}
\int (\dot{S}_D + \dot{S}_i)dM &=& \left( \frac{1}{T_0} - \frac{1}{T_1} \right) F
\\
&=& \frac{F^2\eta(P)}{T_0[T_0+F\eta(P)]}
< \frac{F}{T_0} .
\end{eqnarray*}
Hence, the enstrophy (dissipation) is bounded --supplying us a
constraint while we maximize the energy.  Under the assumption that
$\eta'(P)>0$, maximization of $P$ (energy) raises the entropy
production to its maximum.

The maximum entropy production yields a most ``disordered'' state in
the small-scale, while an ordered flow with the maximum energy appears
in the large-scale of the hierarchy.

Thus the working of this somewhat peculiar ``heat engine'' described
in this paper can be understood in the backdrop of processes and
ideas that have been invoked to study a variety of self-organizing
systems. Special and distinguishing feature of this system is that
when the entering heat flux $F$ exceeds a well-defined threshold, a
transition to a stable state with enhanced temperature gradients
occurs. Simple thermodynamics can capture the essential qualitative
features of the transition as well as of the new state.



\newpage

\vspace*{2cm}



\begin{thebibliography}{11}
\bibitem{N1} it is similar to and reminiscent of statistical mechanics where properties of  macroscopic objects are derived without recourse to the details of molecular forces.
\bibitem{N2} The sign of $\partial F/ \partial T_0$ changes when $T_0 > 4 T_c$. Further increase results in a higher $F_min$. Perhaps the optimal operational region is $T_0 ~ 4 T_c$ where $F_{min}$ has a minimum value.
\end{thebibliography}
\end{document}